\documentclass[aps,prd,showpacs,amssymb,amsmath,nofootinbib]{revtex4}

\usepackage{epsfig,graphicx}

\DeclareMathOperator{\arctanh}{arctanh}
\newcommand{\beq}{\begin{equation}}
\newcommand{\eeq}{\end{equation}}
\newcommand{\bea}{\begin{eqnarray}}
\newcommand{\eea}{\end{eqnarray}}
\newcommand{\bseq}{\begin{subequations}}
\newcommand{\eseq}{\end{subequations}}
\newcommand{\Ref}[1]{(\ref{#1})}

\begin{document}
\title{Exact wormhole solutions with nonminimal kinetic coupling}

\author{R.V. Korolev$^{1}$}
\email{korolyovrv@gmail.com}
\author{S.V. Sushkov$^{2}$}
\email{sergey_sushkov@mail.ru}
\affiliation{Institute of Physics, Kazan Federal University,
Kremlevskaya 18, Kazan 420008, Russia}


\begin{abstract}
We consider static spherically symmetric solutions in the
scalar-tensor theory of gravity with a scalar field possessing the nonminimal
kinetic coupling to the curvature. The lagrangian of the theory contains the
term $(\varepsilon g^{\mu\nu}+\eta G^{\mu\nu})\phi_{,\mu}\phi_{,\nu}$ and
represents a particular case of the general Horndeski lagrangian, which leads to
second-order equations of motion. We use the Rinaldi approach to construct
analytical solutions describing wormholes with nonminimal kinetic coupling. 
It is shown that wormholes exist only if $\varepsilon=-1$ (phantom case) and
$\eta>0$. The wormhole throat connects two anti-de Sitter spacetimes. The
wormhole metric has a coordinate singularity at the throat. However, since all
curvature invariants are regular, there is no curvature singularity there.
\end{abstract}

\maketitle

\section{Introduction}
Natural modifications of general relativity are the models describing possible
nonminimal coupling between matter fields and the curvature. The most
intensively and widely studied are various nonminimal generalizations of
scalar-tensor theories of gravity which have numerous cosmological applications
(see Refs. \cite{SahSta, PeeRat, Nob, CopSamTsu, CalKam, SilTro, Cli_etal,
AmeTsu} for detailed reviews of these and other models).

An especial approach to modified theories of gravity represent models allowing
for nonminimal coupling between derivatives of dynamic quantities of matter
fields and the curvature. The most general scalar-tensor theory of such type was
suggested in the 70-es of the last century in the Horndeski work
\cite{Horndeski}. Horndeski developed his theory on the base of mathematical
facts but later the same results were obtained on the basis of more intuitive
approach from Galileons research \cite{Galileon}.

The simplest Lagrangian in the Horndeski theory contains a term
$G^{\mu\nu}\phi_{,\mu}\phi_{,\nu}$ providing nonminimal kinetic coupling of a
scalar field to the curvature. Cosmological applications of such theory have
been intensively investigated in
\cite{Sus:09,SusSar,Sus:12,SkuSusTop,Cosmology}. In particular, as was shown in
our works \cite{Sus:09,SusSar,Sus:12,SkuSusTop}, nonminimal kinetic coupling
provides essentially new inflation mechanism. Less studied is a problem of black
hole existence in the theory of gravity with nonminimal kinetic coupling. A
while ago Rinaldi \cite{Rinaldi} found a class of exact solutions with
characteristic features of black holes, particularly, with event horizon.
Afterwards, the Rinaldi method was applied in \cite{Ana,Bab,Min,Cis} to find new
particular solutions with event horizons.

In Ref. \cite{SusKor} we have studied wormholes in the theory of gravity with
nonminimal kinetic coupling. Solutions describing asymptotically flat
traversable wormholes have been only obtained by means of numerical methods. The
aim of this work is to construct exact analytical wormhole solutions with
nonminimal kinetic coupling by using the Rinaldi method.

\section{Action and field equations}
Let us consider a gravitational theory with nonminimal derivative
coupling given by the action\footnote{Throughout this paper we use units such that $G=c=1$ and the
conventions for the curvature tensors are $R^\alpha_{\beta\gamma\delta} =
\Gamma^\alpha_{\beta\delta,\gamma} - ...$ and $ R_{\mu\nu} = R^\alpha_{\mu\alpha\nu}$.}
\beq\label{action}
S=\int dx^4\sqrt{-g}\left\{\frac{R}{8\pi} - \left[\varepsilon g_{\mu\nu}+ \eta
G_{\mu\nu}\right]\phi^{,\mu}\phi^{,\nu}\right\},
\eeq
where $g_{\mu\nu}$ is a metric,
$g=\det(g_{\mu\nu})$, $R$ is the scalar curvature, $G_{\mu\nu}$ is the Einstein tensor, $\phi$ is a real massless scalar field, и $\eta$ is a parameter of nonminimal kinetic coupling with the dimension of length-squared.
The $\varepsilon$ parameter equals $\pm1$. In the case $\varepsilon = 1$ we have canonical scalar field with positive kinetic term, and the case $\varepsilon = -1$ describes phantom scalar field with negative kinetic term.

Variation of the action \Ref{action} with respect to the metric $g_{\mu\nu}$ and scalar field
$\phi$ provides the following field equations \cite{Sus:09,SusSar}:
\bseq\label{fieldeq}
\bea
\label{eineq}
&& G_{\mu\nu}=8\pi\big[\varepsilon T_{\mu\nu}
+\eta \Theta_{\mu\nu}\big], \\
\label{eqmo}
&&[\varepsilon g^{\mu\nu}+\eta G^{\mu\nu}]\nabla_{\mu}\nabla_\nu\phi=0,
\eea
\eseq
where
\bea \label{T}
T_{\mu\nu}&=&\nabla_\mu\phi\nabla_\nu\phi-
{\textstyle\frac12}g_{\mu\nu}(\nabla\phi)^2, \\
\Theta_{\mu\nu}&=&-{\textstyle\frac12}\nabla_\mu\phi\,\nabla_\nu\phi\,R
+2\nabla_\alpha\phi\,\nabla_{(\mu}\phi R^\alpha_{\nu)}
\nonumber\\
&&
+\nabla^\alpha\phi\,\nabla^\beta\phi\,R_{\mu\alpha\nu\beta}
+\nabla_\mu\nabla^\alpha\phi\,\nabla_\nu\nabla_\alpha\phi
\nonumber\\
&&
-\nabla_\mu\nabla_\nu\phi\,\square\phi-{\textstyle\frac12}(\nabla\phi)^2
G_{\mu\nu}
\label{Theta}\\
&&
+g_{\mu\nu}\big[-{\textstyle\frac12}\nabla^\alpha\nabla^\beta\phi\,
\nabla_\alpha\nabla_\beta\phi
+{\textstyle\frac12}(\square\phi)^2 -\nabla_\alpha\phi\,\nabla_\beta\phi\,R^{\alpha\beta}
\big]. \nonumber
\eea
Due to the Bianchi identity $\nabla^\mu G_{\mu\nu}=0$, the equation \Ref{eineq}
leads to a differential consequence
\beq
\label{BianchiT}
\nabla^\mu\big[\varepsilon T_{\mu\nu}+\eta \Theta_{\mu\nu}\big]=0.
\eeq
One can check straightforwardly that the substitution of expressions \Ref{T}
and \Ref{Theta} into \Ref{BianchiT}
yields the equation \Ref{eqmo}. In other words, the equation of motion of scalar field \Ref{eqmo} is the differential consequence of the system \Ref{eineq}.

\section{Static spherically symmetric solutions}

Let us find static spherically symmetric solutions of the field equations \Ref{fieldeq}. Under the assumption of spherical symmetry the scalar field is a function of the radial coordinate $r$, i.e. $\phi=\phi(r)$, and the spacetime metric can be taken as follows
\beq \label{genmetric}
ds^2=-f(r)dt^2+g(r)dr^2+\rho^2(r)d\Omega^2,
\eeq
where $ d\Omega^2=d\theta^2+\sin^2\theta d\varphi^2$.
Note that a freedom in choosing the radial coordinate allows us to fix the form of one of the metric functions $f(r)$, $g(r)$ or $\rho(r)$, but at this stage we will not do it for the sake of generality. Now, using the above-mentioned metric and scalar field ansatz, we can represent the field equations \Ref{fieldeq} in the following form:
\begin{subequations} \label{fieldeq1}
\begin{align}
\label{first}
 &\frac{\sqrt{fg}}{g}\psi\left[\varepsilon\rho^2+\eta\left(\frac{\rho\rho'f'}{fg
}+\frac{\rho'^2}{g}-1\right)\right]=C_0,\\
 \label{second}
 &\rho\rho'\frac{f'}{f}=\frac{
g(g-\rho'^2)-4\pi\eta\psi^2(g-3\rho'^2)+4\pi\varepsilon \rho^2\psi^2
g}{g-12\pi\eta\psi^2},\\
 \label{third}
 &\frac{\rho\rho'}{2}\left(\frac{f'}{f}-\frac{g'}{g}\right)=\frac{
g(g-\rho'^2-\rho\rho'')+4\pi\eta\psi^2(2\rho'^2+\rho\rho'')+4\pi\eta
\rho\rho'(\psi^2)'}{ g - 12\pi\eta\psi^2},
\end{align}
\end{subequations}
where $C_0$ is an integration constant, and $\psi\equiv\varphi'$.
It is worth noting that the equation \Ref{first} is a first integral of the
equation of motion \Ref{eqmo}.

Then, following Rinaldi \cite{Rinaldi}, we will search for analytical solutions
of the system \Ref{fieldeq1} in the particular case supposing that
\beq
C_0=0.
\eeq
In this case the equation \Ref{first} yields
\beq\label{first1}
\rho\rho'\frac{f'}{f}=g-\rho'^2-\frac{\varepsilon\rho^2 g}{\eta},
\eeq
This gives the following expression for the function $f(r)$:
\beq\label{genf}
f(r)=\frac{C_1}{\rho}\exp\left(-\int{\frac{(\varepsilon\rho^2-\eta)g}{\eta\rho\rho'}dr}\right),
\eeq
where $C_1$ is an integration constant.
From Eq. \Ref{second}, using the relation \Ref{first1}, one can derive $\psi^2$:
\beq\label{genpsi}
\psi^2(r)=\frac{\varepsilon \rho^2 g}{8\pi\eta(\varepsilon\rho^2-\eta)}.
\eeq

The scalar curvature for the metric \Ref{genmetric} takes the following form:
\beq
R=\frac{2}{\rho^2}-\frac{2\rho' f'}{\rho g f} +\frac{2\rho' g'}{\rho g^2}
-\frac{2\rho'^2}{\rho^2 g}-\frac{4\rho''}{\rho g}+\frac{f'^2}{2gf^2}
-\frac{f''}{g f}+\frac{g'f'}{2g^2 f}.
\eeq
By using the relation \Ref{first1}, one can eliminate the function $f(r)$ from
the expression for $R$. As a result we find
\beq\label{R}
R=\frac{2(\varepsilon\rho^2+\eta)}{\eta\rho^2}
-\frac{3\rho'}{2\rho g}\left(\frac{\rho'}{\rho}-\frac{g'}{g}\right)
-\frac{3\rho''}{\rho g}
-\frac{g}{2\rho^2\rho'^2}
-\frac{\varepsilon g(\varepsilon\rho^2-2\eta)}{2\eta^2\rho'^2}
+\frac{g'(\varepsilon\rho^2-\eta)}{2\eta\rho\rho' g}
-\frac{\rho''(\varepsilon\rho^2-\eta)}{\eta\rho\rho'^2}.
\eeq

Formulas \Ref{genf}, \Ref{genpsi} and \Ref{R} states the functions $f(r)$,
$\psi(r)$ and $R(r)$ in terms of $g(r)$. The equation for $g(r)$ can be obtained
by eliminating $f(r)$ and $\psi(r)$ from equation \Ref{third} by using the
relations \Ref{first1} and \Ref{genpsi}:
\beq\label{geng}
\rho\rho'(\varepsilon\rho^2-2\eta)\frac{g'}{g}-
\left(\textstyle\frac{1}{\eta}\rho^4-3\varepsilon\rho^2 +2\eta\right)g
+\rho'^2\left(3\varepsilon\rho^2+2\eta\right)
-2\rho\rho''(\varepsilon\rho^2-2\eta)-\frac{4\rho^4\rho'^2
} {\varepsilon\rho^2-\eta}=0.
\eeq
It is worth noting that a general solution of Eq. \Ref{geng} could be obtained
analytically for an arbitrary function $\rho(r)$. Depending on the sign of
$\varepsilon\eta$ the solution takes the following forms:

{\bf A. $\varepsilon\eta>0$.}
\bea\label{gA}
g(r) &=& \frac{\rho'^2(\rho^2-2l_{\eta}^2)^2}{(\rho^2-l_{\eta}^2)^2 F(r)},\\
\label{FA}
F(r) &=&
3-\frac{8m}{\rho}-\frac{\rho^2}{3l_\eta^2}+\frac{l_\eta}{\rho}\arctanh\frac{\rho}{l_\eta},
\eea

{\bf B. $\varepsilon\eta<0$.} 
\bea\label{gB}
g(r) &=&\frac{\rho'^2(\rho^2+2l_{\eta}^2)^2}{(\rho^2+l_{\eta}^2)^2 F(r)},\\
\label{FB}
F(r) &=&
3-\frac{8m}{\rho}+\frac{\rho^2}{3l_\eta^2}+\frac{l_\eta}{\rho}\arctan\frac{\rho}{l_\eta},
\eea
Here $m$ is an integration constant and $l_\eta=|\varepsilon\eta|^{1/2}$ is a characteristic scale of nonminimal kinetic coupling.

For the specified function $\rho(r)$ formulas \Ref{gA}-\Ref{FB} together with \Ref{genf} and \Ref{genpsi} give a solution to the problem of $g(r)$, $f(r)$ and $\psi(r)$ construction. Below we consider two special examples of the function $\rho(r)$.

\section{Schwarzschild coordinates: $\rho(r)=r$}
As was mentioned above, a freedom in choosing the radial coordinate $r$ allows to specify additionally the form of one of the metric functions. Let us make a choice:
\beq
\rho(r)=r.
\eeq
This case corresponds to Schwarzschild coordinates, so that $r$ is the curvature radius of coordinate sphere $r={\rm const}>0$.

Substituting $\rho(r)=r$ into the formulas \Ref{genf}, \Ref{genpsi}, \Ref{gA}-\Ref{FB} and calculating the integral in \Ref{genf}, we derive the solutions for $g(r)$, $f(r)$ и $\psi(r)$. For the first time the solutions in the case $\rho(r)=r$ were obtained by Rinaldi \cite{Rinaldi}. Below we briefly discuss these solutions separately depending on a sign of the product $\varepsilon\eta$.

{\bf A. $\varepsilon\eta>0$.}
In this case the solution reads
\bea
 f(r) & = & C_1 F(r), \label{f1}\\
 g(r) & = & \frac{(r^2-2l_\eta^2)^2}{(r^2-l_\eta^2)^2 F(r)}, \label{g1}\\
 \psi^2(r) & = & \frac{\varepsilon}{8\pi l_\eta^2}
 \frac{r^2(r^2-2l_\eta^2)^2}{(r^2-l_\eta^2)^3 F(r)}, \label{psi1}
\eea
where $C_1$ is an integration constant and
\beq
F(r)=3-\frac{r^2}{3 l_\eta^2}-\frac{8m}{r}+\frac{l_\eta}{r}
\arctanh\frac{r}{l_\eta}.
\eeq
In the limit $r\to 0$ the solution \Ref{f1} for the function $f(r)$ takes the asymptotical form
$$
f(r)\approx 4C_1\left(1-\frac{2m}{r}\right).
$$
To compare the derived asymptotic with the Schwarzschild solution it is convenient to set $C_1=\frac14$.

Note that the expressions \Ref{f1}-\Ref{psi1} contain the function $(l_\eta/r)\arctanh r/l_\eta$, which is defined in the domain $r\in(0,l_\eta)$. To continue the solution into the interval $r\in(l_\eta,\infty)$, one should make use of identity
$$
\frac{l_\eta}{r}\arctanh \frac{r}{l_\eta} = \frac{l_\eta}{2r}\ln\frac{l_\eta+r}{l_\eta-r},
$$
and then turn to the function $\frac{l_\eta}{2r}\ln\left|\frac{l_\eta+r}{l_\eta-r}\right|$.
At $r\to\infty$ the asymptotic of the function $f(r)$ with the domain extended to the interval $(l_\eta,\infty)$ has a form of de Sitter solution:
$$
f(r) \approx \frac{3}{4}-\frac{r^2}{12 l_\eta^2}.
$$
Also note that at $r=l_\eta$ the function $\frac{l_\eta}{2r}\ln\left|\frac{l_\eta+r}{l_\eta-r}\right|$ has logarithmic singularity and that is why its domain consists of two disconnected parts ${\cal R}_1: 0<r<l_\eta$ and ${\cal R}_{2}: l_\eta<r<\infty$. This implies that we have two different classes of solutions of the form \Ref{f1}-\Ref{psi1} which are defined independently within separate domains ${\cal R}_1$ and ${\cal R}_2$.


Further let us take into account the fact that we consider the real scalar field, so the value $\psi^2$ should be nonnegative, i.e. $\psi^2\ge0$. In view of this requirement the formula \Ref{psi1} imposes additional restrictions on $r$ domain. In particular, it should be noted that in each of the intervals ${\cal R}_1$ and ${\cal R}_2$ at fixed $\varepsilon$ a sign of the function $\psi^2(r)$ is defined by a sign of $F(r)$ and reverses where the function $F(r)$ reverses its sign. Hence we can resume that the solution \Ref{f1}-\Ref{psi1} cannot be considered as a solution which describes a black hole in the theory of gravity with nonminimal kinetic coupling.

\vskip6pt
{\bf B. $\varepsilon\eta<0$.}
In this case the solution reads as follows
\bea
 f(x) & = & \frac14 F(r), \label{f2}\\
 g(x) & = & \frac{(r^2+2l_\eta^2)^2}{(r^2+l_\eta^2)^2F(r)}, \label{g2}\\
 \psi^2(r) & = & -\frac{\varepsilon}{8\pi l_\eta^2}
 \frac{r^2(r^2+2l_\eta^2)^2}{(r^2+l_\eta^2)^3 F(r)}, \label{psi2}
\eea
where
\beq
F(r)=3+\frac{r^2}{3 l_\eta^2}-\frac{8m}{r}+\frac{l_\eta}{r}
\arctan\frac{r}{l_\eta}.
\eeq
Now the solution contains a function $\frac{l_\eta}{r}\arctan\frac{r}{l_\eta}$ and has a domain $r\in(0,\infty)$.
In the limit $r\to 0$ the function $f(r)$ yields the Schwarzschild asymptotics:
$f(r)\approx 1-\frac{2m}{r}$,
and in the limit $r\to\infty$ -- the anti-de Sitter one:
$f(r) \approx \frac{3}{4}+\frac{r^2}{12 l_\eta^2}.$
However, the obtained solution cannot be considered as an analogue of the Schwarzschild-anti-de Sitter solution, as in the case of $m>0$ the function $F(r)$ reverses sign at a point $r_h$ inside the interval $r\in(0,\infty)$ and hence the value of $\psi^2$ becomes negative in one of the intervals $(0,r_h)$ or $(r_h,\infty)$ according to the sign of $\varepsilon$.

From physical point of view a case $m=0$ may be of some interest. In this case the function $F(r)$ is everywhere positive and regular. About a zero point the metric functions have asymptotics $f(r)=1+O(r^2)$ and $g(r)=1+O(r^2)$, as well as $\psi^2(r)=\frac{\varepsilon}{8\pi l_\eta^2}\frac{r^2}{l_\eta^2}(1+O(r^2))$. Thus, at $\varepsilon=+1$ we obtain static spherically symmetric configuration with regular center and the anti-de Sitter structure at infinity.

\section{Wormhole configuration: $\rho(r)=\sqrt{r^2+a^2}$}
In this section we consider static spherically symmetric configurations with the metric function $\rho(r)$ of the following form:
\beq\label{rhowh}
\rho(r)=\sqrt{r^2+a^2},
\eeq
where $a>0$ is a parameter. Then the metric \Ref{genmetric} reads
\beq\label{whmetric}
ds^2=-f(r)dt^2+g(r)dr^2+(r^2+a^2)d\Omega^2.
\eeq
If $f(r)$ and $g(r)$ are everywhere positive and regular functions with a domain $r\in(-\infty,\infty)$, then the metric \Ref{whmetric} describes a wormhole configuration with a throat at $r=0$; parameter $a$ is a throat radius.

Substituting $\rho(r)=\sqrt{r^2+a^2}$ into the formulas \Ref{genpsi}, \Ref{gA}-\Ref{FB}, we derive the solutions for $g(r)$ and $\psi^2(r)$ in an explicit form. The solution \Ref{genf} for $f(r)$ contains the indefinite integral, which in this case cannot be expressed in terms of elementary functions. Below we consider these solutions for each sign of the product $\varepsilon\eta$.

\vskip6pt
{\bf А. $\varepsilon\eta>0$.}
In this case the solution for $g(r)$ is given by the formulas \Ref{gA}-\Ref{FA}. The solution contains the function $(l_\eta/\sqrt{r^2+a^2})\arctanh (\sqrt{r^2+a^2}/l_\eta)$ with the domain that could be found from condition $\sqrt{r^2+a^2}/l_\eta<1$, i.e. $|r|<r_1\equiv(l_\eta^2-a^2)^{1/2}$. At the points $|r|=r_1$ the function $(l_\eta/\sqrt{r^2+a^2})\arctanh (\sqrt{r^2+a^2}/l_\eta)$ logarithmically diverges. Consequently, the metric function $g(r)$ guides the singular behavior near $|r|=r_1$, that makes this solution inconsiderable from physical point of view.

\vskip6pt
{\bf B. $\varepsilon\eta<0$.}
In this case,  by substituting $\rho(r)=\sqrt{r^2+a^2}$ into the formulas \Ref{gB}-\Ref{FB} and \Ref{genf}, we obtain the following solutions for the metric functions $g(r)$ and $f(r)$ and the function $\psi^2(r)$:
\bea
g(r) &=& \frac{r^2(r^2+a^2+2l_\eta^2)^2}{(r^2+a^2)(r^2+a^2+l_\eta^2)^2 F(r)}, \label{gwh}\\
f(r) &=& \frac{a}{\sqrt{r^2+a^2}}\exp\left[\int_0^r\frac{r(r^2+a^2+2l_\eta^2)^2} {l_\eta^2(r^2+a^2)(r^2+a^2+l_\eta^2)F(r)}dr\right], \label{fwh}\\
\psi^2(r) &=& -\frac{\varepsilon}{8\pi l_\eta^2}
\frac{r^2(r^2+a^2+2l_\eta^2)^2}{(r^2+a^2)(r^2+a^2+l_\eta^2)^3 F(r)}, \label{psiwh}
\eea
where
\beq
F(r)=3-\frac{8m}{\sqrt{r^2+a^2}}+\frac{r^2+a^2}{3l_\eta^2}
+\frac{l_\eta}{\sqrt{r^2+a^2}}\arctan \left(\frac{\sqrt{r^2+a^2}}{l_\eta}\right),
\eeq
and the integration constant $C_1=a$ in the expression for $f(r)$ is chosen so that $f(0)=1$. The function $F(r)$ has a minimum at $r=0$, thus to make it everywhere positive one should demand $F(0)>0$. Hence one can derive the limitation on the upper value of the parameter $m$:
\beq\label{M}
2m<a\left(\frac34+\frac{\alpha^2}{12}+\frac{1}{4\alpha}\arctan\alpha\right),
\eeq
where $\alpha\equiv a/l_\eta$ is the dimensionless parameter which defines the ratio of two characteristic sizes: the wormhole throat radius $a$ and the scale of nonminimal kinetic coupling $l_\eta$. In the particular case $a\ll l_\eta$ we get $2m<a$. Further, we assume that the value of $m$ satisfies the condition \Ref{M}, and therefore the function $F(r)$ is positively definite, i.e. $F(r)>0$.

Let us consider asymptotical properties of the obtained solution. Far from the throat in the limit $|r|\to\infty$ the metric functions $g(r)$ and $f(r)$ have the following asymptotics:
\beq
g(r)=3\frac{l_\eta^2}{r^2}+O\left(\frac{1}{r^4}\right), \quad
f(r)=A\frac{r^2}{l_\eta^2}+O(r^0),
\eeq
where the constant $A$ depends on the parameters $a$, $l_\eta$ and $m$ and can be calculated only numerically. Let us note that the derived asymptotics correspond to geometry of anti-de Sitter space with the constant negative curvature.

In the neighborhood of the throat $r=0$ we obtain
\beq
g(r)=B\frac{r^2}{l_\eta^2}+O(r^4), \quad f(r)=1+O(r^2),
\eeq
where
$$
B=\frac{(\alpha^2+2)^2}{\alpha^2(\alpha^2+1)^2
{\textstyle\left(3+\frac13\alpha^2-\frac{8m}{a}+\frac{1}{\alpha}\arctan\alpha\right)}}.
$$

It is worth noticing that at the throat $r=0$ the metric function $g(r)$
vanishes, i.e. $g(0)=0$.
This implies that there is a coordinate singularity at $r=0$.
To answer the question whether there is a geometric singularity at this point,
one should compute the curvature invariants for the metric \Ref{genmetric}.
In this paper we confine ourselves to discussion of the scalar curvature \Ref{R}.
By substituting the solution \Ref{gwh} into \Ref{R} one can check that the
curvature near the throat is regular: $R(r)=R_0+O(r^2)$, where the value
$R_0=R(0)$ is cumbersomely expressed in terms of the parameters $a$, $l_\eta$ и
$m$.
Far from the throat in the limit $|r|\to\infty$ the scalar curvature tends
asymptotically to a constant negative value, i.e. $R\to R_\infty$, where
\beq
R_\infty=-\frac{5+3l_\eta^2}{2l_\eta^2}.
\eeq
It is worthwhile to note that the asymptotical value $R_\infty$ is determined
only by the characteristic scale of nonminimal kinetic coupling $l_\eta$ and
does not depend on $a$ and $m$. We also note that $R_\infty\to-\infty$ in the
limit $l_\eta\to0$.

Finally, discuss briefly the solution \Ref{psiwh} for $\psi^2(r)$. Since
$F(r)>0$, hence the condition $\psi^2(r)\ge0$ holds only for $\varepsilon=-1$.
Now taking into account that the solutions \Ref{gwh}, \Ref{fwh}, and \Ref{psiwh}
was obtained in the case $\varepsilon \eta<0$, we can conclude that $\eta>0$.

To illustrate the performed analysis, we show numerical solutions for $g(r)$, $f(r)$ and the scalar curvature $R(r)$ in Fig. 1.
\begin{figure}
\includegraphics[width=5cm]{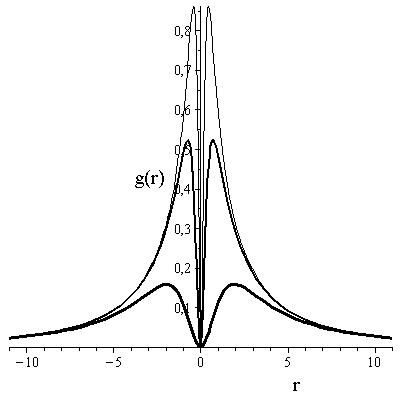} \hfill
\includegraphics[width=5cm]{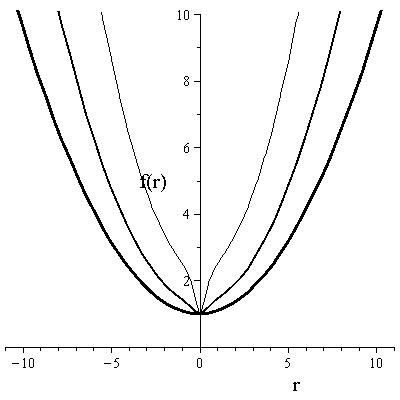} \hfill
\includegraphics[width=5cm]{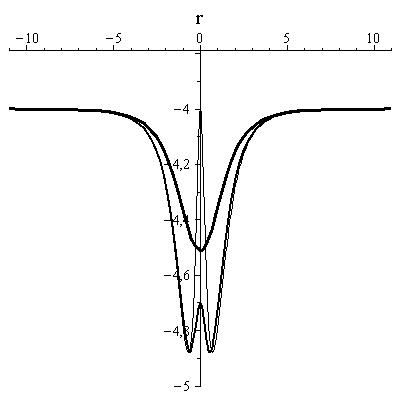}
\caption{Graphs for the metric functions $g(r)$, $f(r)$ and the scalar curvature $R(r)$ with $l_\eta=1$, $m=0.1$. Curves, from thin to thick, are given for  $a=0.3; 0.5; 1.5$.}
\end{figure}

\section{Summary}
In this paper we have explored static spherically symmetric solutions in the
scalar-tensor theory of gravity with a scalar field possessing the nonminimal
kinetic coupling to the curvature. The lagrangian of the theory contains the
term $(\varepsilon g^{\mu\nu}+\eta G^{\mu\nu})\phi_{,\mu}\phi_{,\nu}$ and
represents a particular case of the general Horndeski lagrangian
\cite{Horndeski}, which leads to second-order equations of motion. Previously,
Rinaldi \cite{Rinaldi} found a class of exact solutions with nonminimal kinetic
coupling with characteristic features of black holes, particularly, with event
horizon. In this work, using the Rinaldi approach, we have found and examined
analytical solutions describing wormholes. A  detailed analysis revealed a
number of characteristic features of the obtained solutions. In particular, it
turned out that the wormhole solution exists only if $\varepsilon=-1$ (phantom
case) and $\eta>0$. The wormhole metric has a specific coordinate
singularity at the wormhole throat, namely, the metric component $g_{rr}$ is
vanished at $r=0$. However, there is no curvature singularity at the throat,
since all the curvature invariants stay regular. Also it was shown that the
wormhole throat connects two asymptotical regions with anti-de Sitter geometry
of spacetime.

\section*{Acknowledgments}
The work was supported by the Russian Government Program of Competitive Growth
of Kazan Federal University and by the Russian Foundation for Basic Research
grants No.~14-02-00598. Also R.K. appreciates the Dynasty Foundation for support
and encouragement.

\end{document}